\documentclass[10pt]{article}

\usepackage[margin=2cm]{geometry}
\usepackage{amsfonts}
\usepackage{graphicx} 
\usepackage{subfig}
\usepackage{cite}
\usepackage{caption}
\usepackage{amsmath}
\usepackage{amssymb}
\usepackage{color}
\usepackage{soul}
\usepackage[normalem]{ulem}
\usepackage{authblk}  
\usepackage[utf8]{inputenc}
\usepackage{epsfig}
\usepackage{float} 
\usepackage[toc,page]{appendix}

\usepackage{tikz}
\usetikzlibrary{patterns}
\usetikzlibrary{decorations.text}
\usetikzlibrary{decorations.markings}
\usetikzlibrary{decorations.pathmorphing}

\tikzset{snake it/.style={decorate, decoration=snake}}

\definecolor{EquationGreen}{HTML}{00aa00}
\definecolor{SCO}{HTML}{32CD32}
\definecolor{UCO}{HTML}{FFCC00}
\definecolor{Background}{HTML}{FAFAFA}
\definecolor{FontColor}{HTML}{23373B}
\definecolor{NoCO}{HTML}{8C1A8C}

\definecolor{DarkGreen}{RGB}{0,215,0}

\title{\textbf{EMRIs around $j=1$ black holes with synchronised hair}}
\author[1]{Jorge F. M. Delgado\footnote{jorgedelgado@ua.pt}}
\author[1]{Carlos A. R. Herdeiro\footnote{herdeiro@ua.pt}}
\author[1]{Eugen Radu\footnote{eugen.radu@ua.pt}}

\affil[1]{\normalsize Departamento de Matemática da Universidade de Aveiro and 

Center for Research and Development in Mathematics and Applications -- CIDMA

Campus de Santiago, 3810-183 Aveiro, Portugal}

\begin{document}

\date{April 2023}

\maketitle

\begin{abstract}
\normalsize

We study extreme mass ratio inspirals (EMRIs) due to an infalling Light Compact Object (LCO) onto a generic class of stationary and axi-symmetric massive compact objects (MCO - with or without a horizon). Using the quadrupole hybrid formalism we obtain a master formula for the evolution of the radius of the LCO and find qualitatively different behaviours depending on the geodesic structure of the MCO. We then specialize the MCO to a black hole with synchronised scalar hair (BHsSH). To allow a comparison with a highly spinning Kerr BH, we consider BHsSH with dimensionless spin, $j = 1$. This yields two distinct sequences of solutions. The first harbours Kerr-like solutions with maximal hairiness of $\sim 10\%$. The corresponding EMRIs are Kerr-like, but the cut-off frequency can be a few times smaller than in Kerr, yielding waveforms with quantitatively significant non-Kerrness. The second sequence links the extremal Kerr black hole to a mini-boson star with $j=1$. Here we observe qualitative non-Kerrness, such as the non-monotonically increase of the angular velocity and stagnation endpoints, reflecting Kerr-unlike geodesic structures.

\end{abstract}

\tableofcontents

\newpage

\section{Introduction}

The last few years have proven fruitful for the study of gravitational waves (GWs), powered by the 90 detections of compact binary mergers (between binary black holes (BHs), binary neutron stars, and neutron star--BH pairs) done by the LIGO/Virgo/KAGRA collaboration throughout three scientific runs \cite{LIGOScientific:2018mvr,LIGOScientific:2020ibl,LIGOScientific:2021djp}. These detections opened the window to observe the last moments before the coalescence of compact binary objects, to test General Relativity (GR) in the extreme gravity regime \cite{LIGOScientific:2016lio,LIGOScientific:2018dkp,LIGOScientific:2019fpa,LIGOScientific:2020tif,LIGOScientific:2021sio}, constraining also possible modifications to GR \cite{Yunes:2016jcc,Baker:2017hug,Sakstein:2017xjx,Sennett:2019bpc,Hagihara:2019ihn,Nair:2019iur,Wang:2021jfc,Perkins:2021mhb,Okounkova:2021xjv,Lyu:2022gdr,Fernandes:2022kvg}. 

The great success of the LIGO/Virgo/KAGRA collaboration paved the road to the ongoing implementation of the future Laser Interferometer Space Antenna (LISA). This is a space-based interferometer consisting of 3 spacecraft arranged in an equilateral triangle with an arm-length of 2.5 million kilometers, making it a perfect observatory to measure GWs on the millihertz band. Thus, with LISA, it will be possible to open a different frequency window to new and different astrophysical phenomena hitherto inaccessible to the LIGO/Virgo/KAGRA collaboration \cite{Barausse:2020rsu}. 

One of the new different astrophysical phenomena that will be possible to observe and measure are extreme mass ratio inspirals (EMRIs): inspiral motions of a light compact object (LCO) around a central massive compact object (MCO), orders of magnitude more massive than the former. 
Due to the large mass ratio, the energy lost due to the emission of GWs is a  small fraction of the total energy of the system, leading to a slow evolution where the LCO completes $\sim$tens of thousands of revolutions around the MCO before plunging towards the latter. Therefore, the GWs produced encode a detailed scanning of the MCO spacetime geometry, with a huge potential for testing predictions of GR and modified gravity in the strong gravity regime~\cite{Maselli:2021men,Barsanti:2022vvl,Maselli:2020zgv}. 
Furthermore, also due to the large mass ratio, one does not need to use full numerical relativity methods to evolve the system. Instead, it is possible to use perturbation theory methods, such as solving the Teukolsky equation. This particular method is still rather computational and time-wise expensive and mainly limited to the Kerr spacetime, where separability of the radial and angular dependencies is possible. Nevertheless, advances in solving this equation in more generic spacetimes are being made -- see \textit{eg.} \cite{Maselli:2020zgv,Maggio:2021uge,Luna:2022rql,Li:2022pcy,Hussain:2022ins,Cano:2023tmv}. 

A simpler but informative method to explore new and challenging spacetime geometries is the \textit{hybrid formalism}. This formalism combines the geodesic motion of massive matter around the MCO with  the Einstein's Quadrupole Formula - the lowest order post-Newtonian (pN) dissipative effect. The implementation of this method is simpler, computationally, and time-wise cheaper than the Teukolsky approach and it can be easily computed for a more generic spacetime. 
As such, besides already being applied for Kerr and quasi-Kerr spacetimes \cite{Glampedakis:2002cb,Glampedakis:2002ya,Glampedakis:2005cf,Babak:2006uv,Chua:2017ujo,Destounis:2021mqv}, this formalism was applied for boson stars \cite{Kesden:2004qx}, dynamical-Chern-Simons BHs \cite{Sopuerta:2009iy,Pani:2011xj,Canizares:2012is}, Einstein-scalar-Gauss-Bonnet BHs \cite{Yagi:2012gp}, and, more recently, for BHs with synchronised scalar hair (BHsSH) \cite{Collodel:2021jwi}.

Collodel \textit{et al.}~\cite{Collodel:2021jwi} analysed the EMRIs around three particular solutions of BHsSH. These are solutions of the Einstein-(complex)-Klein-Gordon theory that evades well-known no-hair theorems \cite{Bekenstein:1996pn,Herdeiro:2015gia} by considering a complex scalar field, and were first introduced and computed in \cite{Herdeiro:2014goa}. They are, heuristically, a composite spinning horizon-boson star system. The equilibrium between the scalar field and the BH is attained at the threshold of superradiance, where the angular frequency of the scalar field matches the horizon angular velocity of the BH \cite{Herdeiro:2014goa,Herdeiro:2015gia}. 
Since their introduction, several studies were done to test the viability of these solutions. Some of the studies include their shadows \cite{Cunha:2015yba,Cunha:2016bpi,Cunha:2019ikd}, X-ray phenomenology through the iron K-$\alpha$ line \cite{Ni:2016rhz}, quasi-periodic oscillations \cite{Franchini:2016yvq}, thin accreation disks \cite{Collodel:2021gxu}, polish doughnuts \cite{Gimeno-Soler:2021ifv}, structure of timelike circular orbits \cite{Delgado:2021jxd} and linear perturbations \cite{Ganchev:2017uuo,Degollado:2018ypf}.

The solutions studied in \cite{Collodel:2021jwi} are almost entirely composed of the scalar field, \textit{i.e.} most of the ADM mass of the BH was stored in the scalar field. Moreover they occur in very Kerr-unlike regions of the parameter space, where, for instance, multiple light rings and ergo-regions can exist.  Here, we shall focus on how EMRI features evolve as one transitions from a highly spinning Kerr black hole into a hairy one. For this we fix the dimensionless spin of the analysed solutions, $j \equiv J/M^2 = 1$. This unveils two distinct sequences of solutions. Along one sequence, the black holes always remain fairly Kerr-like, whereas in the other they continuously change into a boson star, becoming very Kerr-unlike. Accordingly, we shall see quantitative only, but potentially phenomenologically significant, changes in the waveforms along the first sequence, wherein, moreover, BHsSH can be formed due to the superradiance instability of bald Kerr BHs~\cite{Herdeiro:2021znw}, further motivating these hairy solutions and their EMRI waveforms. The second sequence, by contrast, offers both quantitative and qualitative differences with respect to Kerr EMRIs, sourced by the Kerr-unlike geodesic structure observed for some of the hairiest solutions.


This work is organised as follows. In section 2, we present the mathematical framework of the hybrid formalism to compute the evolution of EMRIs. We also discuss particular properties of the framework and derive a master formula for the LCO radius evolution. In section 3, we introduce the solutions that we shall study. We give a description and some properties of the solutions and we introduce the two distinct sequences of solutions with fixed dimensionless spin, $j = 1$. We proceed by computing and analysing the EMRIs evolutions for several solutions along both sequences. We conclude this work by presenting some final remarks. We use units with $c=1=G$.

\section{Extreme Mass Ratio Inspirals}

We shall study EMRIs through the quadrupole hybrid formalism, wherein a LCO is in circular motion around a MCO, losing energy by GW emission, modelled by the lowest order pN term, the Einstein's quadrupole formula, and migrating adiabatically into lower circular orbits. Once the last stable circular orbit connected to spatial infinity -- the marginally stable circular orbit (MSCO)\footnote{For Kerr BHs, this is also the innermost stable circular orbit (ISCO), but for a generic compact object this is not always the case.} -- is reached, the LCO plunges rapidly towards the central object.

This approximation ignores all finite size effects of the LCO, such as the spin or tidally induced quadrupole, which are higher order pN effects. These terms may  become relevant near the plunge, depending on the scales involved, but are beyond the scope of the present work. Here, both objects are considered as point-like masses, the central object has mass $M$ and is fixed at the center of our coordinate system,  $(x_\text{C},y_\text{C},z_\text{C})=(0,0,0)$ and the LCO has mass $m$ and it will move in circular orbits on the equatorial plane, such that $(x_\text{LCO}, y_\text{LCO}, z_\text{LCO})=(R \cos \Omega t, R \sin \Omega t, 0)$, where $R$ is an appropriate radial coordinate and $\Omega$ is the angular velocity of the LCO. 
For this setup, the quadrupole moment tensor can be written as~\cite{carroll2019spacetime,Thorne:1980ru}, 
\begin{equation}
	Q_{ij} = \sum_\alpha m_\alpha \left( x_{i\alpha} x_{j\alpha} - \frac{1}{3} \delta_{ij} \mathbf{r}^2 \right) ~,
\end{equation}
where $\alpha = \{1,2\}$ corresponds to the central object and LCO, respectively; $m_\alpha$ are the masses of the objects, $x_{i\alpha} = \{x_\alpha,y_\alpha,z_\alpha\}$ and $\mathbf{r}^2 = x_\alpha^2 + y_\alpha^2 + z_\alpha^2$. Since we fixed the central object to the center of the coordinate system, the quadrupole moment tensor will be written only in terms of the LCO's mass and position.

The quadrupole formula can be easily computed by performing time derivatives of the quadrupole moment tensor~\cite{carroll2019spacetime,Thorne:1980ru},
\begin{equation}\label{Eq:EinsteinQuadrupoleFormula}
	\dot{E} = \frac{1}{5} \sum_{ij}\dddot{Q}_{ij} \dddot{Q}_{ij} ~,
\end{equation}
where the dots denote time derivatives. In the end, for our setup, the energy lost through the emission of GWs will be,
\begin{equation}\label{Eq:QuadrupoleFormulaEMRI}
	\dot{E} = -\frac{32}{5} m^2 R^4 \Omega^6 = -\frac{32}{5} \mu^2 M^2 R^4 \Omega^6 ~,
\end{equation}  
where we introduced the mass ratio between both object, $\mu \equiv m/M$, in the last equation. The power computed through Eq. \eqref{Eq:EinsteinQuadrupoleFormula} is measured at spatial infinity and therefore we introduce a minus sign in Eq. \eqref{Eq:QuadrupoleFormulaEMRI} to represent the loss of energy by the system.

The GW field can be computed through the quadrupole momentum tensor as well as~\cite{carroll2019spacetime,Thorne:1980ru},
\begin{equation}
	h_{ij} = \frac{2}{D} \ddot{Q}_{ij} ~,
\end{equation}
where $D$ is the distance between the LCO and the observer. This field can be decomposed into the two allowed polarization modes yielding,
\begin{equation}\label{Eq:StrainGW}
	h_+ = -\frac{4 \mu M}{D} R^2 \Omega^2 \cos 2 \Omega t \hspace{10pt} , \hspace{10pt} h_\times = -\frac{4 \mu M}{D} R^2 \Omega^2 \sin 2 \Omega t ~.
\end{equation}
Such modes are the strain that gravitational detectors, such as the LIGO/Virgo/KAGRA collaboration and LISA, measure, thus observable quantities that we wish to compute for our model. 

Next, we need information regarding the energy associated with the circular motion of the LCO.  Since we want to study EMRIs for a class of BHs solutions that are stationary, axisymmetric, and asymptotically flat, the metric must reflect such assumptions.
We shall follow~\cite{Delgado:2021jxd} and take a generic metric in the form
\begin{equation}
	ds^2 = g_{tt}(r,\theta) dt^2 + 2g_{t\varphi}(r,\theta) dt d\varphi + g_{\varphi\varphi}(r,\theta) d\varphi^2 + g_{rr}(r,\theta) dr^2 + g_{\theta\theta}(r,\theta) d\theta^2  ~,
\end{equation}
where the coordinates $(t,\varphi)$ were adapted to the Killing vector fields associated with the stationarity and axial symmetries, $\eta_1 = \partial_t$ and $\eta_2 = \partial_\varphi$; the spherical-like coordinates $(r,\theta)$ define a subspace orthogonal to $(t,\varphi)$ due to the imposition of circularity and, by a gauge choice, $(r,\theta)$ are chosen to be orthogonal to each other. Furthermore, a $\mathbb{Z}_2$-symmetry is assumed, 
with the equatorial plane being its fixed point set, located at $\theta = \pi/2$.
 We focus on equatorial, circular motion for the LCO.

To work with a geometrically significant radial coordinate, we introduce, on the equator, the perimetral radius, $r_P = \sqrt{g_{\varphi\varphi}}|_{\theta=\pi/2}$, such that a circumference along the equatorial plane has perimeter $2\pi r_P$. We take this coordinate to play the role of $R$ described previously, $r_P = R$.

With this new radial coordinate, we can rewrite the metric, on the equatorial plane, as,
\begin{equation}
	ds^2 = g_{tt}(R,\pi/2) dt^2 + 2g_{t\varphi}(R,\pi/2) dt d\varphi + R^2 d\varphi^2 + g_{RR}(R,\pi/2) dR^2  ~.
\end{equation}

Dropping (for notation ease) the explicit radial dependence of the metric functions, the LCO motion is ruled by the Lagrangian,
\begin{equation}
	2\mathcal{L} = g_{\mu\nu} \dot{x}^\mu \dot{x}^\nu = g_{tt} \dot{t}^2 + 2 g_{t\varphi} \dot{t} \dot{\varphi} + R^2 \dot{\varphi}^2 + g_{RR} \dot{R}^2 = -1 \ .
\end{equation} 
The two constants of motion associated with the Killing vector fields are the energy, $E/m$, and the angular momentum, $L/m$, both per unit mass of the LCO, given by
\begin{equation}
	-\frac{E}{m} = g_{t\mu} \dot{x}^\mu = g_{tt} \dot{t} + g_{t\varphi} \dot{\varphi} ~, \hspace{10pt} \frac{L}{m} = g_{\varphi\mu} \dot{x}^\mu = g_{t\varphi} \dot{t} + g_{\varphi\varphi} \dot{\varphi} ~.
\end{equation}
The units are consistent with Eq. \eqref{Eq:QuadrupoleFormulaEMRI}.
Using the above equations, we can rewrite the Lagrangian as,
\begin{equation}\label{Eq:Lagrangian_AB}
	2\mathcal{L} = - \frac{A(R,E,L)}{B(R)} + g_{RR} \dot{R}^2  = -1 ~,
\end{equation}
where,
\begin{equation}
	A(R,E,L) \equiv \frac{1}{m^2} \left( R^2 E^2 + 2 g_{t\varphi} E L + g_{tt} L^2 \right) \hspace{10pt} \text{and} \hspace{10pt} B(R) \equiv  g_{t\varphi}^2 - g_{tt} R^2 ~.
\end{equation}
We note that $B(R)$ is always a positive function outside of a possible event horizon. 
One can introduce an effective potential by rewriting Eq. \eqref{Eq:Lagrangian_AB},
\begin{equation}
	V(R) \equiv g_{RR} \dot{R}^2 = -1 + \frac{A(R,E,L)}{B(R)} ~.
\end{equation}
A circular orbit can be obtained by solving the following two equations simultaneously,
\begin{eqnarray}
	V(R_\text{cir}) = 0 \hspace{10pt} \Leftrightarrow \hspace{10pt} A(R_\text{cir},E,L) = B(R_\text{cir}) ~, \label{Eq:FirstEqPotential}\\
	V'(R_\text{cir}) = 0 \hspace{10pt} \Leftrightarrow \hspace{10pt} A'(R_\text{cir},E,L) = B'(R_\text{cir}) ~, \label{Eq:SecondEqPotential}
\end{eqnarray}
where prime denotes radial derivative, and we used the first equation to obtain the second one. 
To solve these equations, it proves fruitfully to introduce the angular velocity of the LCO, 
\begin{equation}
	\Omega = \frac{d\varphi}{dt} = \frac{\dot{\varphi}}{\dot{t}} = - \frac{E g_{t\varphi} + L g_{tt}}{E R^2 + L g_{t\varphi}} ~.
\end{equation}
We can use it to solve Eq. \eqref{Eq:FirstEqPotential} and find the energy and angular momentum of the LCO written in terms of the metric functions and the angular velocity,
\begin{equation}\label{Eq:EnergyAngMom}
	\frac{E_\pm}{m} =  \left. - \frac{g_{tt} + \Omega_\pm g_{t\varphi}}{\sqrt{\beta_\pm}} \right|_{r_\text{cir}} ~, \hspace{10pt} \frac{L_\pm}{m} = \left. \frac{g_{t\varphi} + R^2 \Omega_\pm}{\sqrt{\beta_\pm}} \right|_{r_\text{cir}} ~,
\end{equation}
where 
$$\beta_\pm \equiv -g_{tt} - 2 g_{t\varphi} \Omega_\pm - R^2 \Omega_\pm^2 \ .
$$
At the radial coordinate in which $\beta_\pm = 0$ a light-ring is present \cite{Delgado:2021jxd,Delgado:2022yvg}. Then,  Eq. \eqref{Eq:SecondEqPotential} gives an algebraic equation for the angular velocity written in terms of the metric functions,
\begin{equation}\label{Eq:AngVelTCOs}
	\Omega_\pm = \frac{-g_{t\varphi}' \pm \sqrt{C(R)}}{2R} ~,
\end{equation} 
where 
$$
C(R) \equiv g_{t\varphi}'^2 - 2 g_{tt}' R \ ,
$$ 
and $\pm$ correspond to prograde ($+$) and retrograde ($-$) orbits. 
Note that  in order to have any type of circular motion (timelike, null, or spacelike), the function $C(R)$ must be non-negative~\cite{Delgado:2021jxd}. Since the LCO will follow an adiabatic sequence of timelike circular orbits, $C(R) > 0$ for any value of the perimetral radius, $R$, throughout the inspiral.

We can also study the radial stability of the orbit by looking at the sign of the second derivative of the effective potential, which reads, upon the use of Eqs. \eqref{Eq:FirstEqPotential} and \eqref{Eq:SecondEqPotential},
\begin{equation}\label{Eq:StabilityTCOs}
	V''(R_\text{cir}) = \frac{A''(R_\text{cir},E,L) + B''(R_\text{cir})}{B(R_\text{cir})} ~.
\end{equation}
A negative (positive) sign implies a stable (unstable) circular orbit.

We can now combine the emission formula, Eq. \eqref{Eq:QuadrupoleFormulaEMRI}, with the energy of circular orbits. The variation of the orbital energy as a variation of the radial coordinate reads
\begin{equation}
	\dot{E_\pm} = \frac{\partial E_\pm}{\partial R} \dot{R} ~. 
\end{equation}
Computing $dE_\pm/dR$ through the left equation in \eqref{Eq:EnergyAngMom}, it is possible to show that~\cite{Lehebel:2022yyz},
\begin{equation}
	\frac{\partial E_\pm}{\partial R} = \mp \frac{m}{2} \frac{B \Omega_\pm V''}{\sqrt{\beta_\pm C}} \ .
\end{equation}
Thus, the variation of the radial coordinate of the LCO due to the emission of GWs will be
\begin{equation}\label{Eq:RadialEvolution}
	\dot{R} = \pm \frac{64}{5} \mu M R^4 \Omega_\pm^5 \frac{\sqrt{\beta_\pm C}}{B V''} \ .
\end{equation}

From this \textit{master formula} for the radial evolution of a LCO driven by quadrupole GW emission, under the stated assumptions, we can unpack several features:
\begin{enumerate}
\item[$\bullet$]  far away from the central object, the motion should be Keplerian, the orbits being stable, and thus $V'' < 0$. For prograde orbits, the remaining quantities in Eq. \eqref{Eq:RadialEvolution} are all positive, thus $\dot{R} < 0$. For retrograde orbits, all remaining quantities are positive, except the angular velocity, $\Omega_- < 0$, thus, $\dot{R} < 0$ as well. Consequently, the LCO will always inspiral inwards, as one would expect;
\item[$\bullet$] as the LCO inspirals inwards, it may reach a MSCO, where $V'' = 0$. At this threshold, $\dot{R} \rightarrow - \infty$. This result is unphysical signaling a breakdown of the quadrupole hybrid formalism. But the trend in the neighbourhood of the threshold is informative. As the LCO approaches the threshold, $V''$ becomes increasingly smaller and $\dot{R}$ increasingly larger, leading to a faster inspiral. Furthermore, the divergence of $\dot{R}$ can be interpreted as the plunge of the LCO towards the central object;
\item[$\bullet$]
it is possible to have $\dot{R} = 0$.
This can happen when $\beta_\pm = 0$, which, as mentioned before corresponds to a light-ring. The LCO, however, will never reach the light-ring because the outermost light-ring will be always unstable \cite{Cunha:2017qtt,Cunha:2020azh} and that implies the existence of a region of unstable circular orbits radially above the light-ring~~\cite{Delgado:2021jxd,Delgado:2022yvg}. Thus, the LCO will reach the region of unstable circular orbits before the light-ring; 
\item[$\bullet$]	
a different way to have $\dot{R} = 0$ is when $C = 0$, at the threshold between regions with, $C > 0$, and without, $C < 0$, timelike circular orbits. If the region with timelike circular orbits is stable, then it is possible for the LCO to inspiral until it reaches the region without timelike circular orbits -- this can happen, for example, for several classes of boson stars~\cite{Delgado:2021jxd,Delgado:2020udb}. If that happens, $\dot{R} \rightarrow 0$ as the LCO approaches the threshold $C = 0$, leading to a slower and slower inspiral, and it will take an infinite amount of time for the LCO to reach the threshold; 
\item[$\bullet$]		
the final way to have $\dot{R} = 0$ is when $\Omega_\pm = 0$. This is only possible if $g'_{tt} = 0$, and it depends on the sign of $g'_{t\varphi}$. If $g'_{t\varphi} > 0$, only the prograde angular velocity will vanish, $\Omega_+ = 0$, when $g'_{tt} = 0$. On the other hand, if $g'_{t\varphi} < 0$, only the retrograde one will vanish, $\Omega_- = 0$. These points correspond to static rings (SRs)~\cite{Collodel:2017end,Teodoro:2020kok,Teodoro:2021ezj,Collodel:2021jwi,Lehebel:2022yyz}.
Similarly to the previous case, if a SR is connected to stable timelike circular orbits, the LCO can inspiral towards the ring, but it will never reach it since $\dot{R} \rightarrow 0$.
\end{enumerate}

In our modelling, the adiabatic LCO migration towards lower stable circular orbits ends when it reaches:
\begin{description}
\item[i)] either a region of unstable circular orbits, where $V''(R_\text{cir}) > 0$;
\item[ii)]  or a region where one does not have timelike circular orbits, $C(R) < 0$;
\item[iii)] or a SR, $\Omega_\pm = 0$.
\end{description}

An interesting consequence of the existence of SRs in a spacetime is the fact that they may act as an attractor for infalling matter~\cite{Lehebel:2022yyz}. To see this, let us start by computing the stability of a timelike circular orbit coincident with the SR,
\begin{equation}
	V''(R_\text{SR}) = -\left[ \frac{g_{tt}''}{g_{tt}} + \frac{2 (g_{t\varphi}')^2}{B} \right]_{R_\text{SR}}~.
\end{equation}
One observes that the second term is always positive. 
The sign of the first term, however, is more subtle. Its denominator is always negative, $g_{tt} < 0$, because the energy of such TCO is given by $E_\pm = \sqrt{-g_{tt}}$, thus $g_{tt}$ must always be negative at a SR~\footnote{This result also implies that a SR for timelike particles will never form within an ergoregion, where $g_{tt} > 0$.}. Therefore, the sign of the first term depends only on $g_{tt}''$. 
If the SR is localised at a maximum of $g_{tt}$, such that $g_{tt}'' < 0$, then $V''(R_\text{SR})$ will always be negative, and, thus the SR will lay within a region of stable TCOs.
If the SR is localised at a minimum of $g_{tt}$, such that $g_{tt}'' > 0$, no general conclusion is possible, the specific value depending on the relative magnitude between the two terms. Thus, this type of SR can lay within a region of stable or unstable TCOs.

For simplicity, let us consider the case where the SR is located at a maximum of $g_{tt}$. Taylor expanding \eqref{Eq:RadialEvolution} close to the SR, we obtain,
\begin{equation}
	\dot{R} =  \pm \frac{64}{5} \mu M R_\text{SR}^4  \left[(\Omega_\pm')^5 \frac{\sqrt{\beta_\pm C}}{B V''} \right]_{R_\text{SR}}  (R - R_\text{SR})^5 + \mathcal{O} (R-R_\text{SR})^6 ~.
\end{equation}
Computing the derivative of the angular velocity at the SR, we arrive at,
\begin{equation}
	\Omega_\pm'(R_\text{SR}) = \left[ - \frac{g_{tt}''}{2 g_{t\varphi}'} \right]_{R_\text{SR}} ~.
\end{equation}
We have already fixed the sign of $g_{tt}''$ since we have chosen  a SR at a maximum of $g_{tt}$; thus $g_{tt}'' < 0$. The sign of $g_{t\varphi}'$ will depend if we are considering prograde or retrograde orbits:
\begin{itemize}
	\item for prograde orbits, a SR will develop if $g_{t\varphi}' > 0$ on the SR, therefore $\Omega_\pm'(R_\text{SR}) = \dfrac{1}{2} \left| \dfrac{g_{tt}''}{g_{t\varphi}'} \right| > 0$;
	
	\item for retrograde orbits, a SR will develop if $g_{t\varphi}' < 0$ on the SR, therefore $\Omega_\pm'(R_\text{SR}) = - \dfrac{1}{2} \left| \dfrac{g_{tt}''}{g_{t\varphi}'} \right| < 0$.
\end{itemize}
Hence,
\begin{equation}
	\dot{R} =  \frac{2}{5} \mu M R_\text{SR}^4  \left[\frac{\sqrt{\beta_\pm C}}{B V''} \left| \frac{g_{tt}''}{g_{t\varphi}'} \right|^5 \right]_{R_\text{SR}}  (R - R_\text{SR})^5 + \mathcal{O} (R-R_\text{SR})^6 ~.
\end{equation}
This result shows that, for both rotating senses, since we are in a region with stable TCOs, $V''(R_\text{SR}) < 0$, 
\begin{itemize}
	\item if we consider a circular orbit radially above the static ring, $R > R_\text{SR}$, the variation of its radius will decrease, $\dot{R} < 0$, therefore, the particle will \textit{inspiral} towards the SR;
	
	\item if we consider a circular orbit radially below the static ring, $R < R_\text{SR}$, the variation of its radius will increase, $\dot{R} > 0$, therefore, the particle will \textit{outspiral} towards the SR.
\end{itemize}
Therefore, the existence of a maximum of $g_{tt}$ leads to the existence of a static ring within a region of stable TCOs that will act as an attractor for infalling matter for both rotating senses.
However, similarly to the previous case when $C = 0$, it will take an infinite amount of time for the infalling matter to inspiral/outspiral towards the SR, unless some sort of friction acts.

\begin{figure}[h!]
	\centering
	\begin{tikzpicture}
		\filldraw[green!10] circle (2.5) node[black,font=\bfseries] at (0,-1.7){Inspiral};
		\filldraw[red!10,draw=orange, dotted, ultra thick] circle (1.3) node[black,font=\bfseries] at (0,0.8){Outspiral} ;
		\filldraw[black!50!white] circle (0.5);
		\draw [black, thick, postaction={decoration={markings, mark= between positions 0.1 and 0.9 step 2cm with {\arrowreversed{latex}}},decorate},  domain=12*2*pi:7.43*6*pi, samples=400] plot ({\x*cos(2*2*\x)/2/pi/9)}, {\x*sin(2*2*\x)/2/pi/9} );
		\draw [black, thick, postaction={decoration={markings, mark= between positions 0.1 and 0.9 step 1cm with {\arrow{latex}}},decorate},  domain=6*2*pi:12*2*pi, samples=400] plot ({\x*cos(2*2*\x)/2/pi/9)}, {\x*sin(2*2*\x)/2/pi/9} );
		\draw[decoration={text along path, text={|\bf|Static Ring}, text align={center}, text color={orange}, raise=0.1cm, reverse path},decorate] circle (1.3) ;
		\filldraw[black] (-2.3,-0.9) circle (0.08);
		\filldraw[black] (0.55,-1.2) circle (0.08);
	\end{tikzpicture}
	\caption{Illustration of the attractor behaviour of a static ring located at a maximum of $g_{tt}$.}
\end{figure}
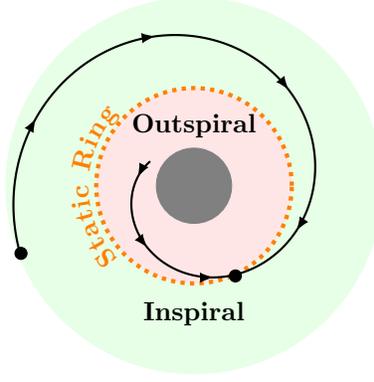

We shall now apply this formalism to a concrete family of non-Kerr black hole spacetimes.

%
%
%
%
%
%
%

\section{Black Holes with Synchronised Scalar Hair}

BHs with synchronised scalar hair~\cite{Herdeiro:2014goa} (BHsSH) -- see also~\cite{Herdeiro:2015gia} -- are 4-dimensional, regular on and outside of the event horizon, stationary, axisymmetric and asymptotically flat solutions of the (complex-)Einstein-Klein-Gordon model, where a complex scalar field, $\Psi$, is minimally coupled to Einstein's gravity. The action of such a model reads,
\begin{equation}
	\mathcal{S} = \int d^4 x \sqrt{-g} \left[ \frac{\mathcal{R}}{16\pi} - g^{\alpha\beta} \partial_\alpha \Psi^* \partial_\beta \Psi - m_\mu^2 \Psi^* \Psi \right]~,
\label{action}
\end{equation}
where $\mathcal{R}$ is the Ricci scalar, $m_\mu$ is the mass of the scalar field and the asterisk ($^*$) denotes the complex conjugate of the scalar field.

The existence of these BHs resides in a crucial condition that stipulates a synchronisation between the angular frequency, $\omega$, and the angular velocity of the horizon of the BH, $\Omega_H$ must exist. This is known as the \textit{synchronisation condition}~\cite{Herdeiro:2014ima}, and it can be written as $\omega = \bar{m} \Omega_H$, where $\bar{m}$ is the azimuthal harmonic index of the scalar field. 
Failure to fulfill this condition leads to the evolution of the system, where the scalar field will either decay towards the BH (if $\omega > \bar{m} \Omega_H$) or be amplified by the BH through a superradiance process (if $\omega < \bar{m} \Omega_H$).

This family of solutions was first obtained in~\cite{Herdeiro:2014goa}, where the authors introduced the novel solutions and studied some basic properties and phenomenology. The solutions were obtained numerically through the use of the CADSOL package~\cite{solver1,solver2,solver3} and the use of suitable ansatz for the metric and scalar field. Such ansatz was,
\begin{equation}
	ds^2 = -e^{2F_0(r,\theta)} \mathcal{N} dt^2 + e^{2F_1(r,\theta)} \left( \frac{dr^2}{\mathcal{N}} + r^2 d\theta^2 \right) + e^{2F_2(r,\theta)} r^2 \sin^2 \theta \left[ d \varphi - W(r,\theta) dt \right]^2~, \hspace{10pt} \mathcal{N} \equiv 1 - \frac{r_H}{r}~,
\end{equation}
\begin{equation}
	\Psi = \phi(r,\theta)\ e^{i(\bar{m} \varphi - \omega t)}~,
\end{equation}
where $\mathcal{F}=\{F_1, F_2, F_0, W, \phi\}$ only depend on the $(r,\theta)$ coordinates and $r_H$ is the radial coordinate of the event horizon.

Fixing $\bar{m}=1$, and considering the fundamental family of spinning solutions, the hairy BHs are described by three global charges: the ADM mass, $M$, the total angular momentum, $J$, and a Noether charge encapsulating the amount of scalar matter. 
The ADM mass and angular momentum are computed through the sum of two contributions, $M = M_H + M_\Psi$ and $J = J_H + J_\Psi$. The first (horizon) contribution, $M_H$ and $J_H$, is computed through the Komar integral~\cite{Poisson:2009pwt,Townsend:1997ku} by taking a spatial section of the horizon, $\mathcal{H}$, as the integrating surface. The second (bulk) contribution, $M_\Psi$ and $J_\Psi$, is computed through the following volume integral\cite{Poisson:2009pwt},
\begin{equation}
	M_\Psi = - 2 \int_\Sigma dS_\mu \left( T_\nu^\mu \xi^\nu - \frac{1}{2} T \xi^\mu \right)~, \hspace{10pt} J_\Psi = \int_\Sigma dS_\mu \left( T_\nu^\mu \eta^\nu - \frac{1}{2} T \eta^\mu \right)~,
\end{equation}
where $\Sigma$ is the spacelike surface bounded by the 2-sphere at infinity, $S_\infty^2$, and the spatial section of the horizon, $\mathcal{H}$; $\xi$ and $\eta$ are the Killing vector fields associated to stationarity and axisymmetry and $T_\nu^\mu$ is the energy-momentum tensor associated to the scalar field.
The Noether charge, $Q$, arises from the global $U(1)$ symmetry of the scalar field. In particular, the transformation 
$\Psi \rightarrow e^{i\alpha}\Psi$, 
where $\alpha$ is a constant, leaves the action unchanged, meaning a conserved current and charge exist,
\begin{equation}
	j^\alpha = -i \left( \Psi^* \partial^\alpha \Psi - \Psi \partial^\alpha \Psi^* \right)~, \hspace{10pt} Q = \int dr d\theta d\varphi \sqrt{-g} j^t~.
\end{equation}
The Noether charge is not entirely independent, but it is related to the angular momentum part of the scalar field, $J_\Psi$, as $J_\Psi = m Q$~\cite{Herdeiro:2014goa,Herdeiro:2015gia}.

The family of BHsSH continuously connects Kerr BHs to mini-boson stars (BSs) ~\cite{Schunck:1996he,Yoshida:1997qf}. The former are the well-known solutions of the vacuum Einstein field equations, where no scalar field exists. Thus, for these bald solutions, $M = M_H$ and $J = J_H$. The latter correspond to horizonless, everywhere regular solitonic solutions, entirely composed of the scalar field. Hence, for these solitonic solutions, $M = M_\Psi$ and $J = J_\Psi$.

We can define the following quantity,
\begin{equation}
	p = 	\frac{M_\Psi}{M} ~,
\end{equation}
known as \textit{hairiness} to quantify how hairy a given solution is. If $p = 0$, the solution is a Kerr BH; if $p = 1$, the solution is a BS.

\begin{figure}[h!]
	\centering
	\includegraphics[scale=0.75]{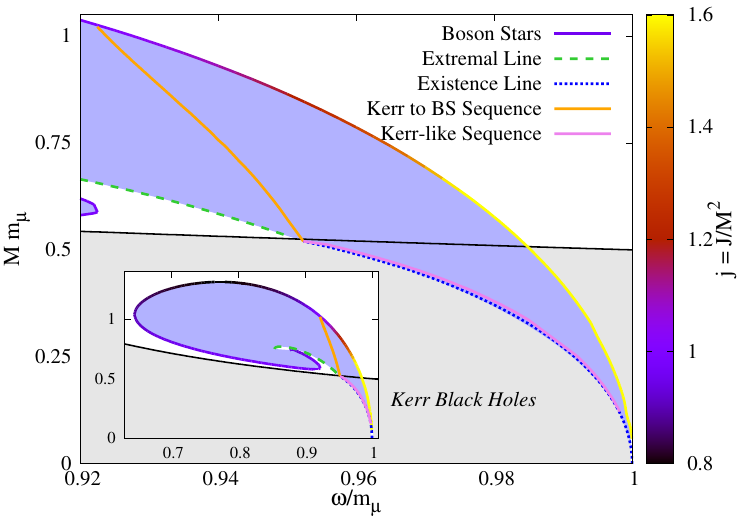}
	\caption{Domain of existence of BHsSH (inset) and a zoom in a particular mass-frequency range (main panel). Note the sequences of hairy solutions with $j=1$.  The orange line corresponds to BHsSH that go from the extremal Kerr solution to the BS line, thus $0 \leq p \leq 1$. The violet line correspond to BHsSH that are Kerr-like, where $p < p_\text{max} \sim 0.1$. This particular sequence lies close to the \textit{Existence Line} (in dotted blue). The \textit{Boson Stars} line has a color gradient corresponding to the value of their dimensionless spin, $j = j/M^2$. We remark that the maximal value of $j$ reported in the colorbar was chosen for representation purposes. BSs can have larger spins close to $\omega/m_\mu = 1$ -- see bottom panels of Fig. 6 in \cite{Delgado:2020udb}.
	}
	\label{Fig:DomainExistence}
\end{figure}

In Fig. \ref{Fig:DomainExistence}, we show part of the domain of existence of BHsSH in a mass vs. frequency diagram, both in units of $m_\mu$. The domain is bounded by three lines. The (solid gradient) BSs line, where one only finds BSs and $p=1$; the (dotted blue) existence line, where one finds Kerr BHs that can support scalar clouds and $p = 0$; the (dashed green) extremal hairy line, where the Hawking temperature vanishes.
We also show another two (solid orange and pink) lines, corresponding to two sequences of hairy solutions with dimensionless spin $j \equiv j/M^2 = 1$. These will be the focus of our analysis, in the following.

\begin{figure}[h!]
	\centering
	\includegraphics[scale=0.65]{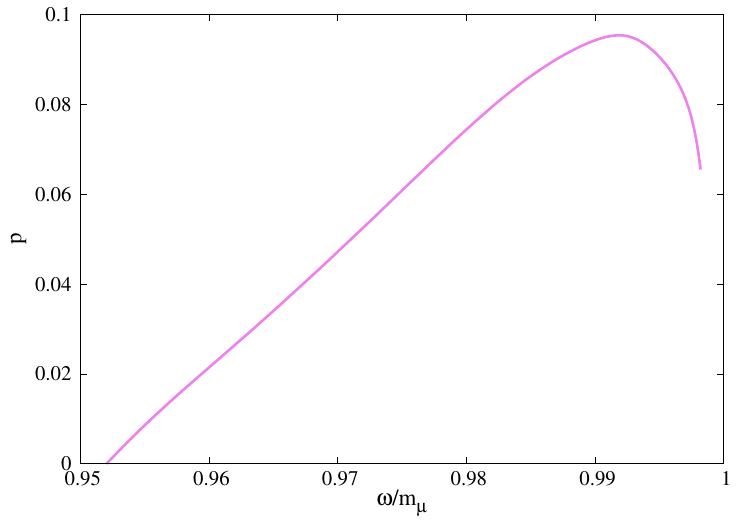}
	\includegraphics[scale=0.65]{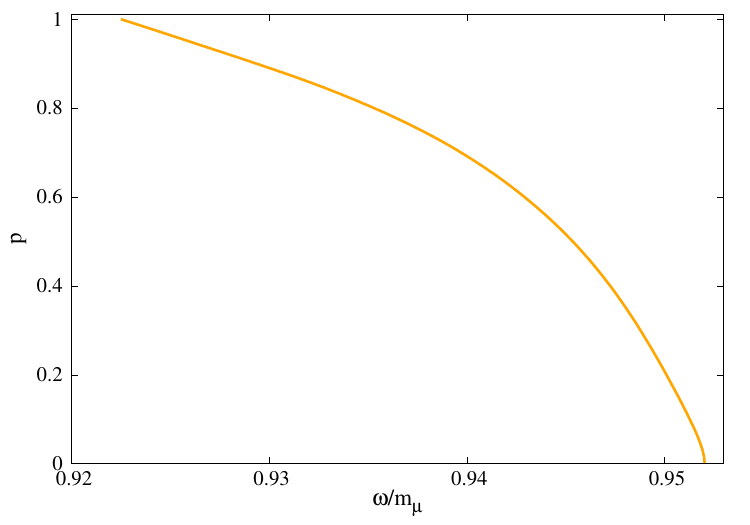}
	\caption{Hairiness of BHsSH with $j=1$ on the Kerr-like sequence (left plot) and on the Kerr to BS sequence (right plot).}
	\label{Fig:HairinessSolutions}
\end{figure}

One section of the $j=1$ hairy BHs sequence (the orange line) starts on the BS with $j=1$. The mass $M$ (angular frequency $\omega$) of the solutions decreases (increseas) until the extremal Kerr BH sitting on the existence line is reached. Along this sequence, the hairiness decreases monotonically from $p=1$ to $p=0$ - right panel of Fig. \ref{Fig:HairinessSolutions}. This is the \textit{Kerr to BS sequence}. 

The other section of the $j=1$ hairy BHs sequence (the violet line), starts at the same extremal Kerr BH on the existence line. Again, the mass $M$ (angular frequency $\omega$) of the solutions decreases (increseas), but now approaching $\omega/\mu = 1$ and $M\mu = 0$, a limit which is numerically inaccessible. Here, the hairiness of the solutions does not exceed $p_\text{max} \sim 0.1$ -- \textit{cf.}~\cite{Herdeiro:2021znw} - 
 left panel of Fig. \ref{Fig:HairinessSolutions}. 
This is the \textit{Kerr-like sequence}.


In the following we shall consider EMRIs where the MCO is a solution along this $j=1$ sequence. It is of interest to compute some physical quantities at the last stable circular orbit that is connected to spatial infinity -- the MSCO, which may, or not, be the ISCO --, where the simulation of the EMRI ends, in order to compare with the Kerr case.

The setup for the evolution is always as follows. We start the simulation with the LCO at $R/M = 10$ and solve numerically Eq. \eqref{Eq:RadialEvolution} through a Runge-Kutta method. We let the simulation evolve until we reach either the MSCO, $V'' = 0$, or a large enough time, that we set to $t/M = 10^{13}$.

\subsection{Kerr-like Sequence}

In this sequence the hairy BHs are fairly Kerr-like, with maximum hairiness of $\sim$10\% - left plot of Fig.~\ref{Fig:HairinessSolutions}. One thus expects no dramatic differences with respect to the Kerr EMRIs. 

\begin{figure}[h!]
	\centering
	\includegraphics[scale=0.65]{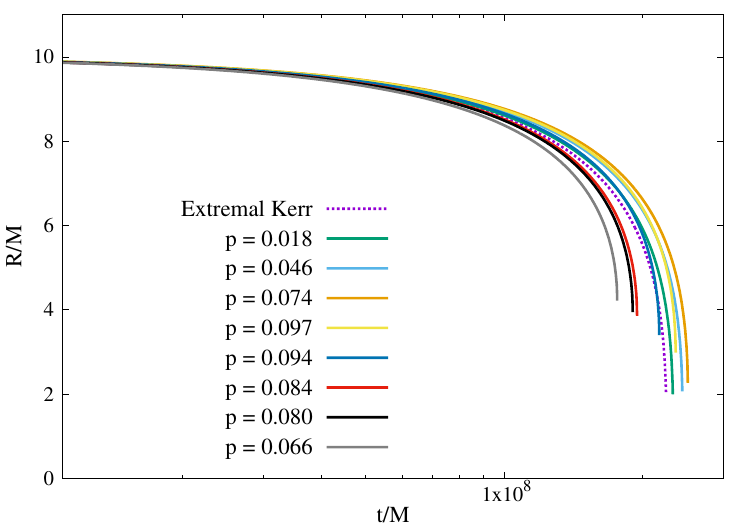}
	\includegraphics[scale=0.65]{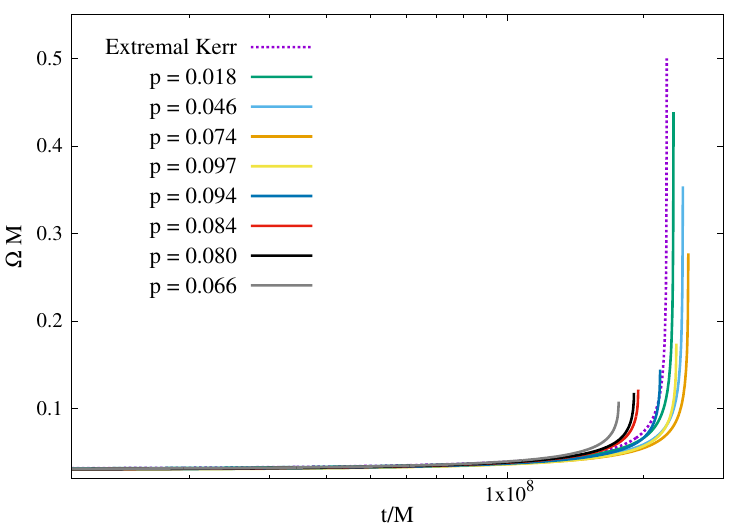}
	\caption{Evolution of the perimetral radius, $R/M$, (left panel) and angular velocity, $\Omega M$, (right panel) of the EMRI for a set of  solutions along the Kerr-like sequence (solid lines) including an extremal Kerr BH (dotted line).}
	\label{Fig:EvolutionEMRIKerrLikeLine}
\end{figure}

\begin{figure}[h!]
	\centering
	\includegraphics[scale=0.65]{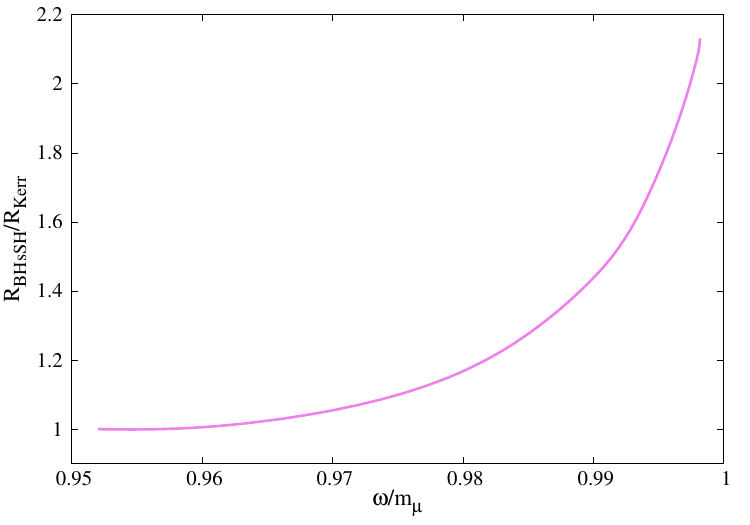}
	\includegraphics[scale=0.65]{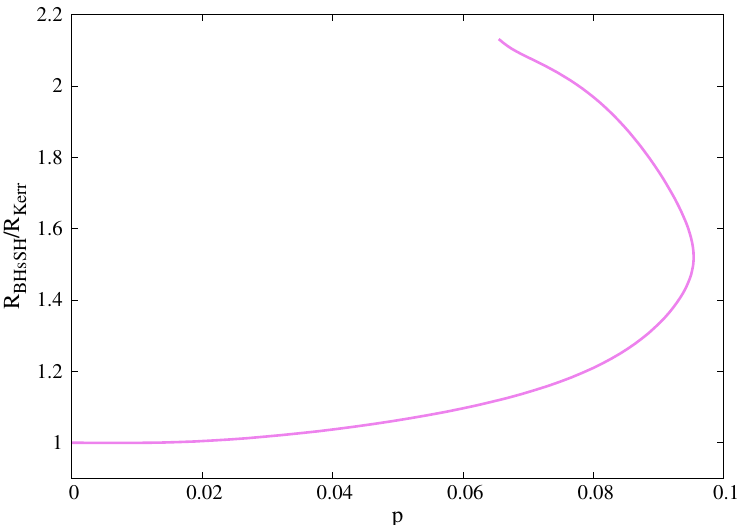}
	\caption{Ratio of the perimetral radius computed at the MSCO between Kerr-like hairy solutions and Kerr BHs. In the left panel we show the dependence with the angular frequency, $\omega/m_\mu$. In the right panel, we show the dependence with the hairiness, $p$.}
	\label{Fig:RatioR_KerrLike}
\end{figure}

\begin{figure}[h!]
	\centering
	\includegraphics[scale=0.65]{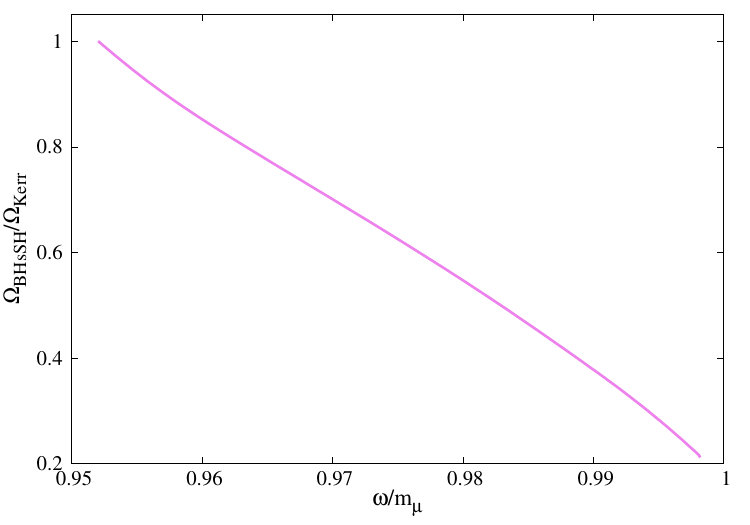}
	\includegraphics[scale=0.65]{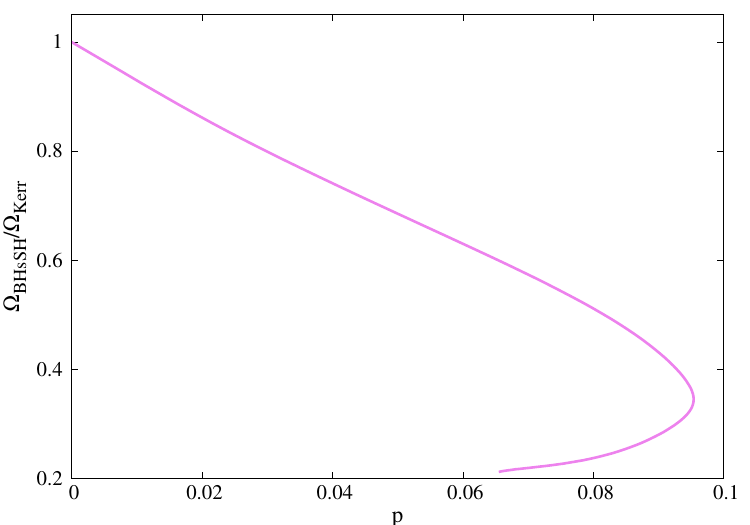}
	\caption{Ratio of the perimetral radius computed at the MSCO between Kerr-like hairy solutions and Kerr BHs. Left panel: dependence with the angular frequency, $\omega/\mu$. Right panel: dependence with the hairiness, $p$. }
	\label{Fig:RatioOmega_KerrLike}
\end{figure}

Fig. \ref{Fig:EvolutionEMRIKerrLikeLine}, shows the perimetral radius, $R/M$, (left plot) and angular velocity, $\Omega$, (in units of $M$, right plot), evolutions for a set of illustrative solutions along this sequence, including an extremal Kerr BH. 
The evolution of the hairy solutions is not dramatically different from the extremal Kerr one. The LCO inspiral is slow, at the start, but monotonically accelerating, until it reaches the MSCO, which, in all cases here is also the ISCO. 
A closer examination reveals that, at the endpoint, $R$ and $\Omega$ can  differ significantly from the extremal Kerr case. For the latter, $R/M = 2$ and $\Omega M = 1/2$, respectively. For the former, we show in Fig. \ref{Fig:RatioR_KerrLike} the ratio between the endpoint value of $R$ for the hairy solutions and extremal Kerr, $R_\text{BHsSH}/R_\text{Kerr}$ as a function of the angular frequency, $\omega/m_\mu$, (left panel) and the hairiness, $p$, (right panel). Likewise for the latter, we present a similar set of panels for the angular velocity ratios, $\Omega_\text{BHsSH}/\Omega_\text{Kerr}$ - Fig.~\ref{Fig:RatioOmega_KerrLike}.


Fig. \ref{Fig:RatioR_KerrLike} (left panel) informs that the ratio between the endpoint $R$ of the hairy solutions and the extremal Kerr BH grows monotonically as $\omega/m_\mu$, increases. At the Kerr limit, where $\omega/m_\mu = \Omega_H/m_\mu = 0.952$, the ratio obviously converges to unity. As $\omega/m_\mu$ increases, the ratio increases until  the last hairy solution considered, $\omega/m_\mu = 0.9982$ and $R_\text{BHsSH}/R_\text{Kerr} \sim 2.13$.  Thus, the endpoint $R$ for this hairy solution can be more than twice the extremal Kerr one. 
Interestingly, the largest deviation from Kerr does not correlate directly with the largest hairiness. Fig. \ref{Fig:RatioR_KerrLike} (right panel) shows  that the hairiest solution (with $p \sim 0.097$)  has only a ratio of $R_\text{BHsSH}/R_\text{Kerr} \sim 1.50$.

The same type of analysis can be done for the $\Omega$ endpoint value, which translates directly into the cut-off frequency of the produced GWs.  Fig. \ref{Fig:RatioOmega_KerrLike} (left panel), shows the ratio between the endpoint value of $\Omega$ for the hairy and extremal Kerr BHs, as a function of $\omega/m_\mu$. The ratio decreases monotonically (and almost linearly) with $\omega/m_\mu$, reaching $\Omega_\text{KBHsSH}/\Omega_\text{Kerr} \sim 0.21$ for the largest $\omega/m_\mu$ studied. For this particular solution, the cut-off frequency of the produced GWs will be $\sim$ a fifth of the corresponding extremal Kerr one.  
Again, the solution with the largest ratio deviation is not the hairiest - \ref{Fig:RatioOmega_KerrLike} (right panel); for the latter $\Omega_\text{BHsSH}/\Omega_\text{Kerr} \sim 0.34$.

\begin{figure}[h!]
	\centering
	\includegraphics[scale=0.7]{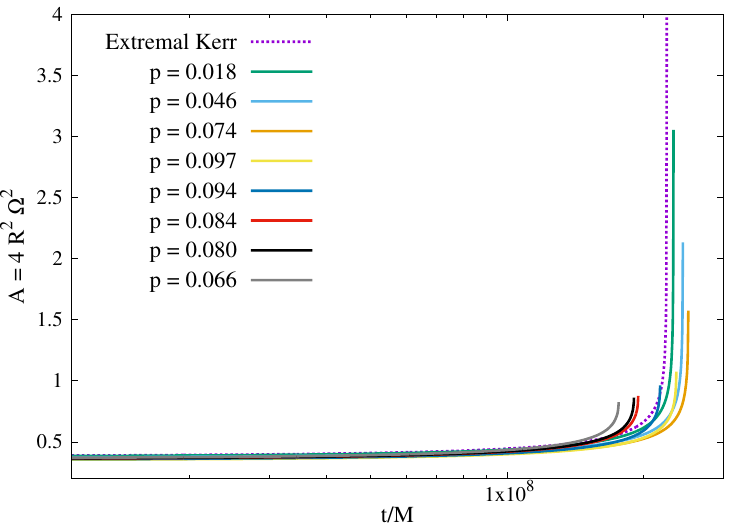}
	\caption{Amplitude of  GWs produced by an EMRI for a set of hairy solutions within the Kerr-like sequence (solid lines), including an extremal Kerr BH (dotted line).}
	\label{Fig:AmplitudeGWKerrLikeLine}
\end{figure}

\begin{figure}[h!]
	\centering
	\includegraphics[scale=0.65]{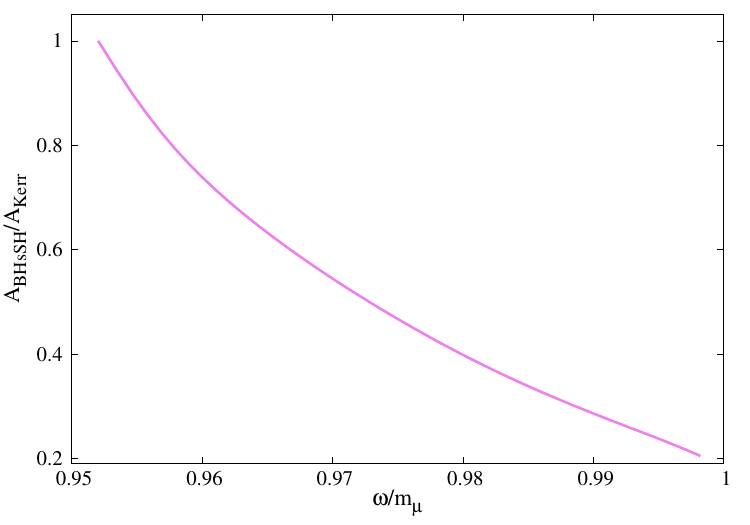}
	\includegraphics[scale=0.65]{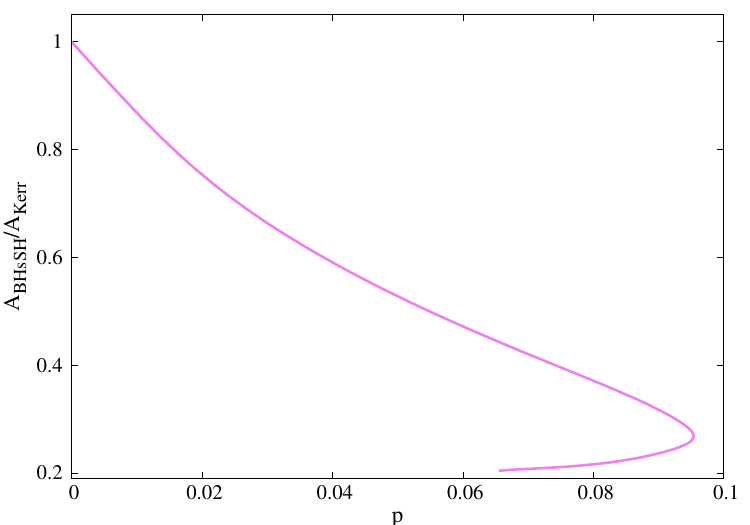}
	\caption{Ratio of the strain computed at the MSCO between Kerr-like hairy solutions and Kerr BHs. Left panel: dependence with the angular frequency, $\omega/\mu$. Right panel: dependency with the hairiness, $p$.  }
	\label{Fig:Ratiohp_KerrLike}
\end{figure}

The previous analysis provides the GW frequency in our approximation, \textit{cf.} Eq. \eqref{Eq:StrainGW}, which is twice the angular velocity, $f_\text{GW} = 2 \Omega$. To get waveforms for the  EMRI we also need the strain $h_+$ amplitude (which is the same as $h_\times$) -- \textit{cf.} Eq. \eqref{Eq:StrainGW} --,
\begin{equation}
	\frac{h_+ D}{\mu M} = 4 R^2 \Omega^2 \equiv A ~.
\end{equation}
Fig. \ref{Fig:AmplitudeGWKerrLikeLine}, shows the evolution of this amplitude.
For all Kerr-like solutions, the strain amplitude evolves similarly to that of the extremal Kerr BH, but the maximal value is maximized by the latter; hair decreases the maximal amplitude of the strain. This is detailed in Fig. \ref{Fig:Ratiohp_KerrLike}, showing the ratio between the maximal value of the strain for hairy solutions and of the extremal Kerr BH, as a function of $\omega/m_\mu$ (left panel) and of the hairiness, $p$ (right panel). The solution with the smallest ratio is the one with the largest $\omega/m_\mu$ studied here, with $A \sim 0.20$ and $\omega/\mu = 0.9982$. Note the similarity with  the angular velocity -- \textit{cf.} left panel of Fig. \ref{Fig:RatioOmega_KerrLike}.  The amplitude trend results from a competition between the $R$ and $\Omega$ trends, $cf.$ Eq.~\eqref{Eq:StrainGW}.   Since the ratio of the $\Omega$ endpoint decreases more than the $R$ endpoint ratio increases (see Figs. \ref{Fig:RatioR_KerrLike} and \ref{Fig:RatioOmega_KerrLike}), the  strain ratio  follows the $\Omega$ trend. Once more, the strain difference is not maximised for the hairiest solution - Fig. \ref{Fig:Ratiohp_KerrLike} (right panel) - but for the solution with $p \sim 0.066$.

\begin{figure}[h!]
	\centering
	\includegraphics[scale=0.65]{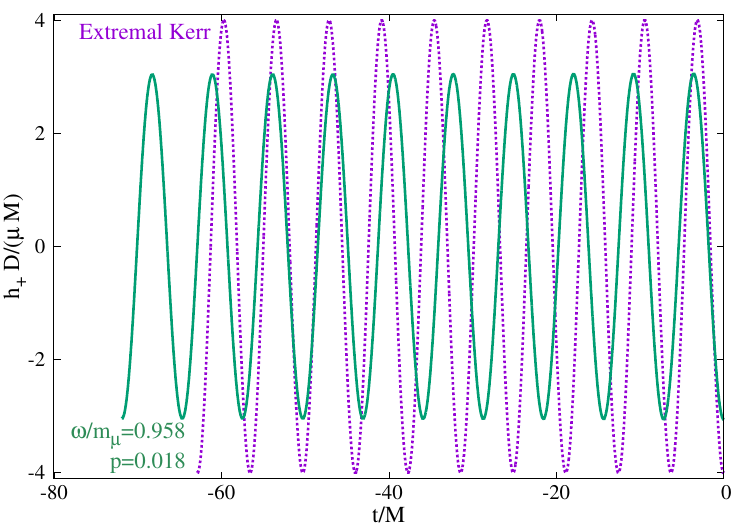}
	\includegraphics[scale=0.65]{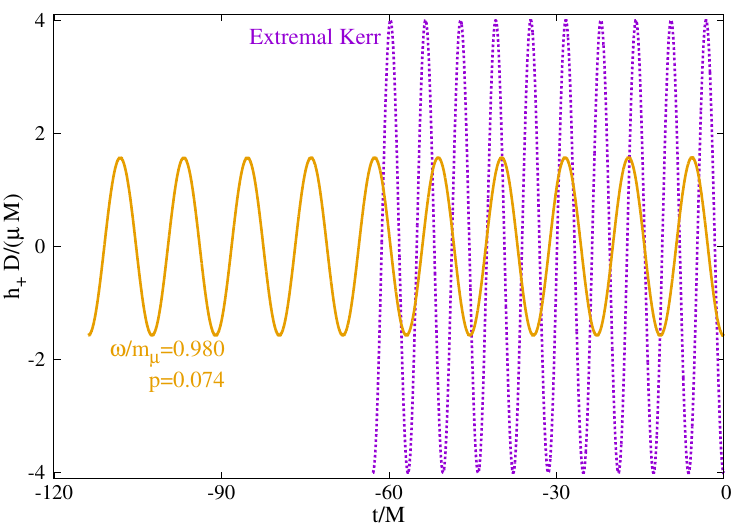}
	\includegraphics[scale=0.65]{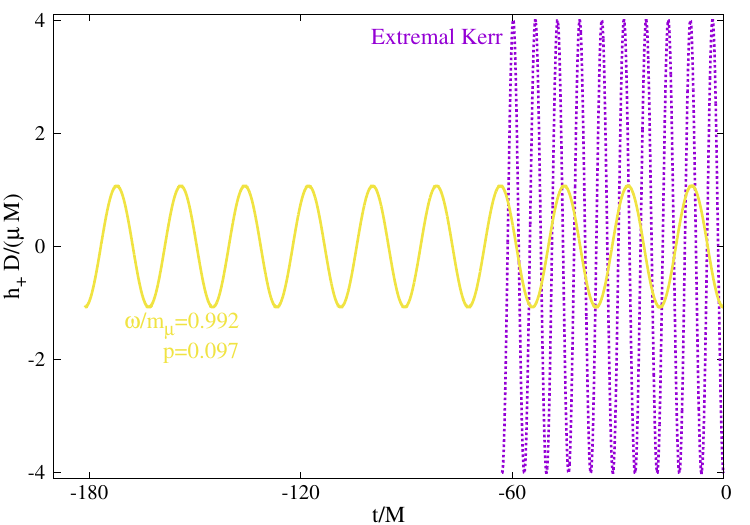}
	\includegraphics[scale=0.65]{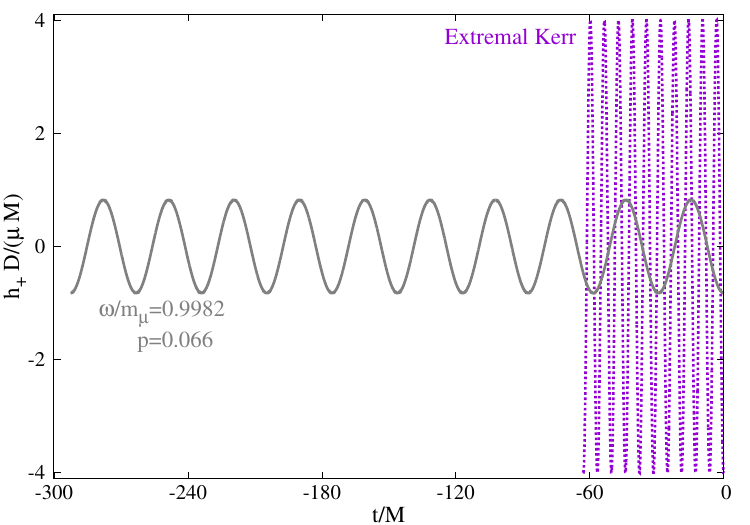}
	\caption{Waveforms from the last 5 revolutions of an EMRI for 4 particular cases of BHsSH in the Kerr-like sequence. The extremal Kerr waveform is also shown (dotted line). 
The same colour scheme as in Figs. \ref{Fig:EvolutionEMRIKerrLikeLine} and \ref{Fig:AmplitudeGWKerrLikeLine} was used. }
	\label{Fig:GWs_KerrLike}
\end{figure}

To conclude, $\lesssim$ 10\% of hair can change the endpoint value of the perimetral radius by a factor of $\lesssim 2$ and of the angular velocity and maximal strain by a factor of around $\gtrsim 1/5$.
Combining the previous phase plus amplitude information, we can illustrate the EMRI waveforms. Fig. \ref{Fig:GWs_KerrLike} shows waveforms generated from the last 5 revolutions of the LCO around 4 particular cases of BHsSH within the Kerr-like sequence, together with the corresponding extremal Kerr waveforms. The chosen hairy solutions possess the following angular frequencies and hairiness: $\omega/m_\mu = \{0.958,0.980,0.992,0.9982\}$ and $p=\{0.018,0.074,0.097,0.066\}$.  Inspection of Fig. \ref{Fig:GWs_KerrLike} produces a strong case that  the GWs produced by the BHsSH EMRI  can be easily distinguished by LISA. Quantitative predictions require, however, an analysis accounting for higher pN terms, in particular quantifying the magnitude of finite-size effects.

\subsection{Kerr to BS Sequence}

We now turn to solutions with $j=1$ connecting the extremal Kerr BH, where no hair exists, $p = 0$, to the rotating mini-boson star with $j=1$, where no event horizon exists, $p = 1$. Here, one may expect qualitative (and not only quantitative) deviation from the Kerr results. Indeed, this is confirmed by the analysis below, which,  however, must be interpreted with care, since important effects, such as, \textit{e.g.} the drag caused by the scalar field \cite{Cardoso:2022whc}, are being neglected. Nevertheless, the analysis and results presented in this subsection provide an idea of how the scalar field can influence EMRIs. 

\begin{figure}[h!]
	\centering
	\includegraphics[scale=0.65]{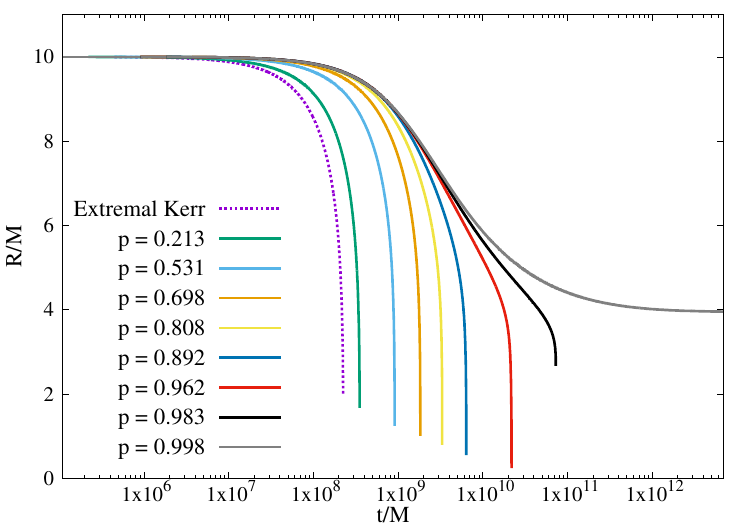}
	\includegraphics[scale=0.65]{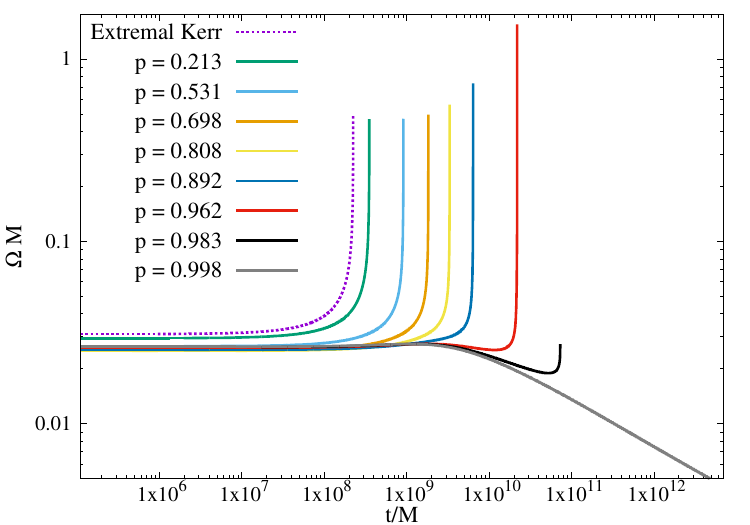}
	\caption{Evolution of the perimetral radius, $R/M$, (left panel) and angular velocity, $\Omega M$, (right panel) of the EMRI for  solutions within the Kerr to BSs sequence (solid lines), including an extremal Kerr BH (dotted line).}
	\label{Fig:EvolutionEMRIKerrToBSsLine}
\end{figure}

Fig. \ref{Fig:EvolutionEMRIKerrToBSsLine} shows the evolution of the perimetral radius, $R/M$ (left panel) and the angular velocity, $\Omega M$ (right panel) of the LCO. 
One now observes both quantitative and qualitative differences in the evolutions of $R$ and $\Omega$ with respect to the extremal Kerr ones. As one adds hair (increasing $p$), initially the inspiral is slower. The MSCO (which, in these cases, is the ISCO) is located at smaller perimetral radii and the $\Omega$ endpoint becomes slightly smaller than the extremal Kerr one. But for $0.65 \lesssim p \lesssim 0.94$, the $\Omega$ endpoint becomes larger than the extremal Kerr one. Furthermore, for $ p \gtrsim 0.94$, the EMRI evolution starts to be vastly different.

\begin{figure}[h!]
	\centering
	\includegraphics[scale=0.65]{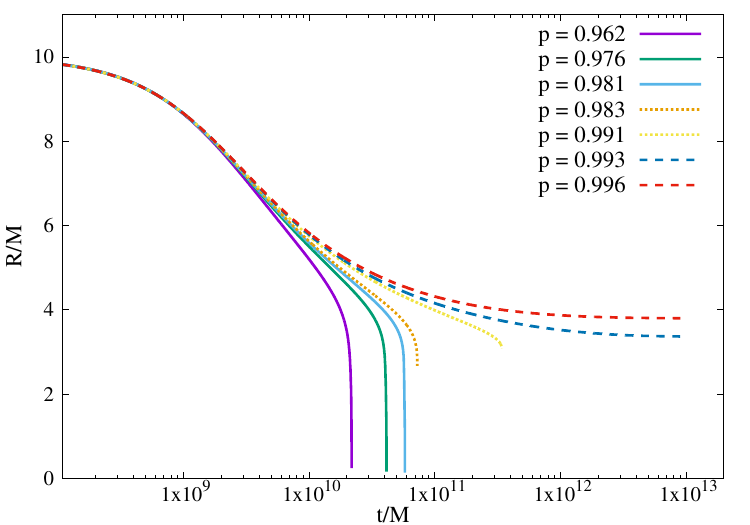}
	\includegraphics[scale=0.65]{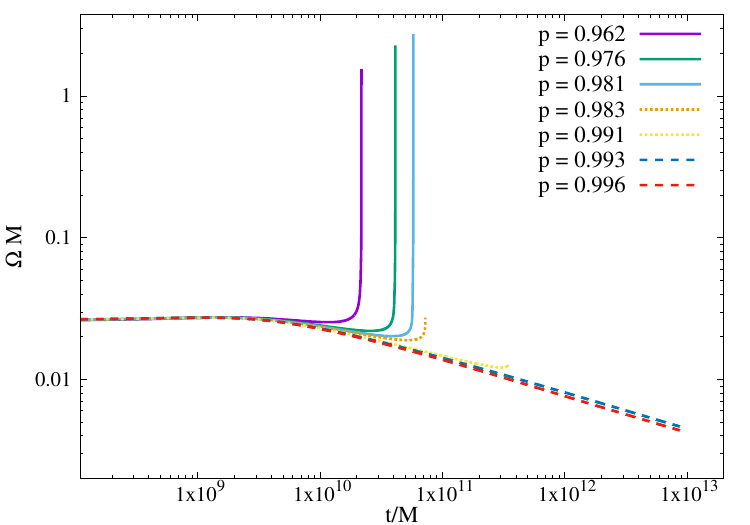}
	\caption{Evolution of the perimetral radius, $R/M$, (left panel) and angular velocity, $\Omega M$, (right panel) of the EMRI for a set of very hairy solutions, $p \gtrsim 0.94$, within the Kerr to BSs sequence. We distinguish the three types of evolutions by the three style of lines. Solid line correspond to evolutions up to the ISCO. Dotted lines correspond to evolutions up to the MSCO, where $V''(R_\text{MSCO}) = 0$. Dashed lines correspond to evolutions up to the MSCO, where $C(R_\text{MSCO}) = 0$.}
	\label{Fig:EvolutionEMRICloseToBSsLine}
\end{figure}

\begin{figure}[h!]
	\centering
	\includegraphics[scale=0.6]{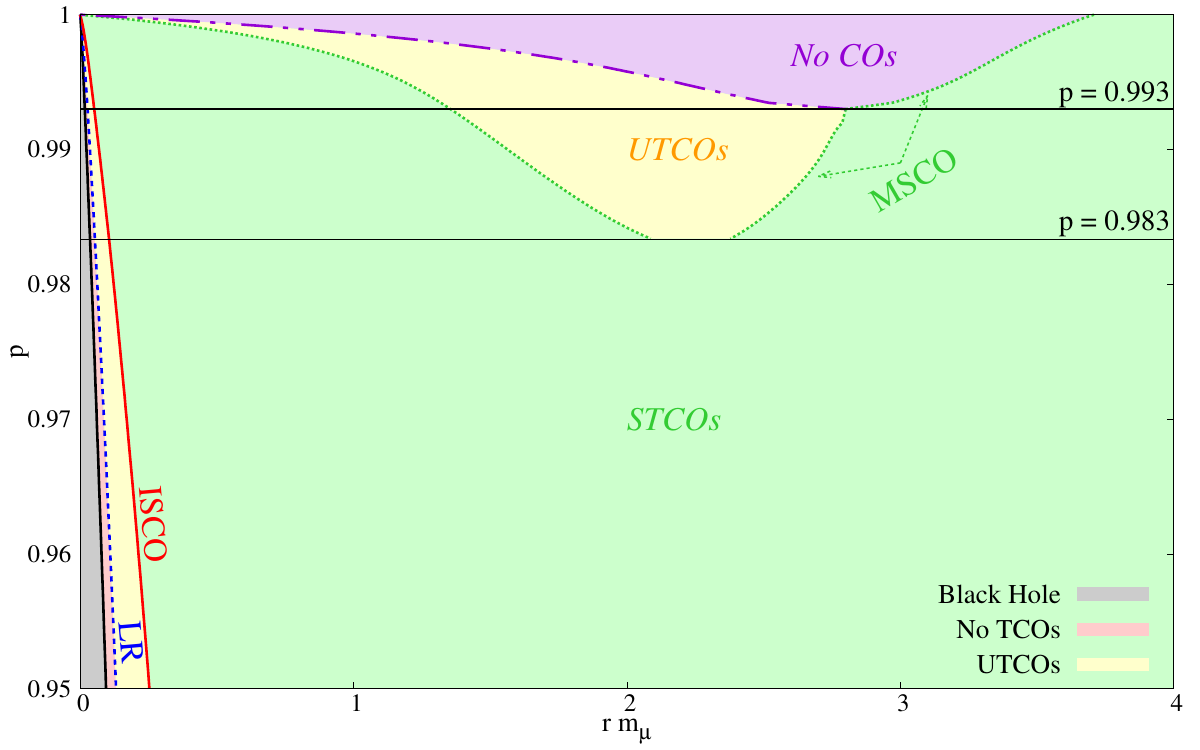}
	\caption{Structure of circular orbits of very hairy ($p > 0.95$) BHsSH in the Kerr to BS sequence.  The green region corresponds to stable timelike circular orbits (STCOs).  }
	\label{Fig:StructureCircularOrbits_KerrToBSLine}
\end{figure}

Let us momentarily focus on the latter and consider Fig. \ref{Fig:EvolutionEMRICloseToBSsLine}.
Firstly, for solutions with $0.94 \lesssim p \lesssim 0.983$, $\Omega$ no longer increases monotonically throughout the evolution. After an initial increase, it decreases slightly before a final chirp as the LCO approaches the ISCO - see $e.g.$ the case with $p = 0.976$ (green solid line in Fig. \ref{Fig:EvolutionEMRICloseToBSsLine}). 
This may be explained by the off-centered distribution of the scalar field energy density -- toroidal in shape --  which, in these cases, contain between $94\%$ and $98.3\%$ of the total spacetime energy. This is consistent with the analysis performed in \cite{Collodel:2021jwi} (for a different region of the parameter space).
Then, for even hairier solutions, $0.983 \lesssim p \lesssim 0.993$, the endpoint of the evolution changes drastically. The $R$ endpoint increases whereas the $\Omega$ endpoint  decreases by orders of magnitude. This is due to the existence of a new region of unstable timelike circular orbits (UTCOs), as can be observed in Fig. \ref{Fig:StructureCircularOrbits_KerrToBSLine}, exhibiting the structure of circular orbits in a hairiness, $p$, \textit{vs} metric radial coordinate, $r m_\mu$ plot. Here, any horizontal line corresponds to a single BHsSH with $j=1$. Several regions are presented. The grey region is within the horizon; the pink region corresponds to a region where no timelike circular orbits exist (No TCOs); the yellow region harbours UTCOs, whereas the green region harbours stable timelike circular orbits (STCOs); finally, the violet region corresponds to a region where no type of circular orbits exist (No COs).
For solutions in the aforementioned range of hairiness, $0.983 \lesssim p \lesssim 0.993$, there is an ISCO (red solid line) and a MSCO (green dotted line). The evolution of the EMRIs for these solutions will stop at the MSCO, which is always situated further away from the BH than the ISCO. This explains the large jump of the $R$ endpoint, and consequently of the angular velocity.

Finally, for solutions with $p \gtrsim 0.993$, the evolution of the EMRI stagnates and no fast inspiral is observed as in the previous cases. This is due to the existence of the violet No COs region. Here, the function $C(R)$ that appears in Eq. \eqref{Eq:AngVelTCOs}, becomes negative, yielding complex angular velocities. This means that, at this new region's boundary, $C(R) = 0$, the variation of the perimetral radius goes to zero, $\dot{R} = 0$, thus stalling the evolution and explaining the results in Fig. \ref{Fig:EvolutionEMRICloseToBSsLine}. This is a concrete realization of the general discussion in Section 2.

\begin{figure}[h!]
	\centering
	\includegraphics[scale=0.65]{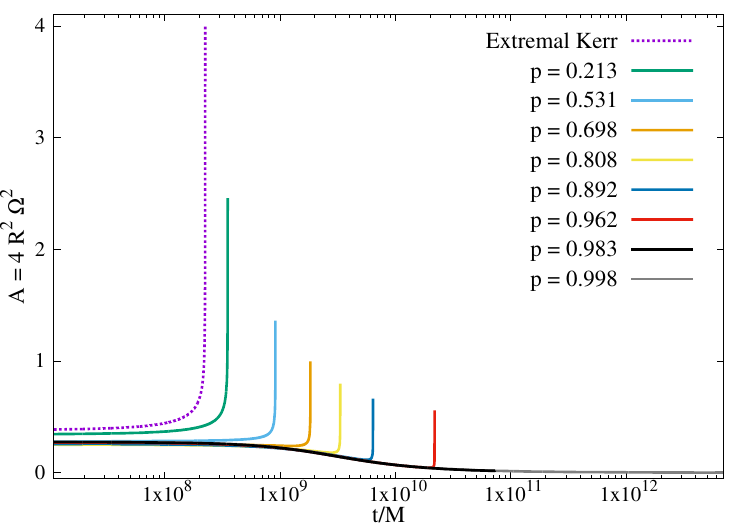}
	\includegraphics[scale=0.65]{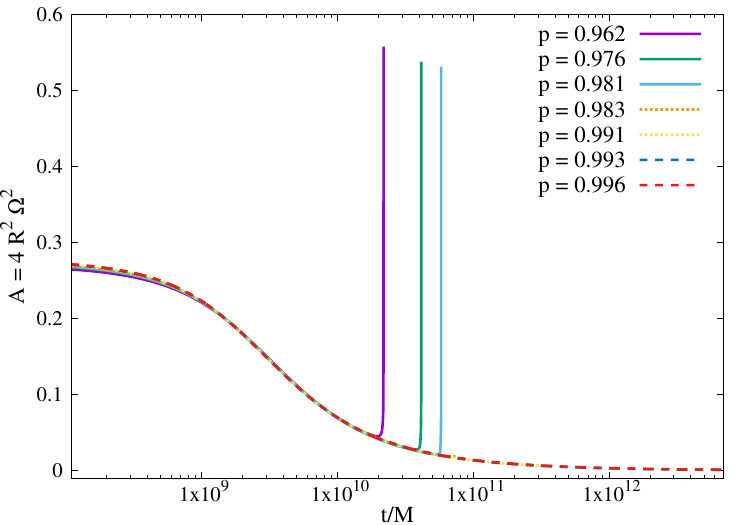}
	\caption{Amplitude of the generated GW produced by an EMRI for a set of hairy solutions (left panel) and very hairy solutions (right panel) within the Kerr to BS sequence. On the left panel we also show the amplitude for the extremal Kerr case as a dotted line. On the right panel we distinguish three types of evolutions by the three types of line, similarly to Fig. \ref{Fig:EvolutionEMRICloseToBSsLine}.}
	\label{Fig:AmplitudeGWKerrToBSsLine}
\end{figure}

\begin{figure}[h!]
	\centering
	\includegraphics[scale=0.7]{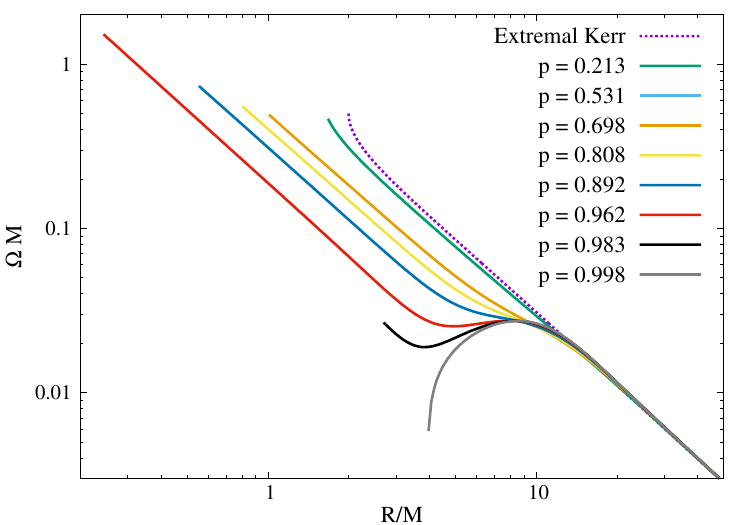}
	\caption{Angular velocity of the LCO \textit{versus} the perimetral radius for a set of hairy solutions within the Kerr to BS sequence. We also show the angular velocity for the extremal Kerr case as a dotted line.} 
	\label{Fig:AngVelKerrToBSsLine}
\end{figure}

Let us now turn to the EMRI waveforms in the Kerr to BS sequence. Fig. \ref{Fig:AmplitudeGWKerrToBSsLine} shows the amplitude of the GWs produced in the cases of the hairy solutions (left panel) and the very hairy solutions (right panel). 
The left panel informs that increasing the amount of hair  decreases the overall amplitude of the produced GWs, mainly at the end of the evolution. 
This can be explained by studying $\Omega$ as a function of $R$, shown in Fig. \ref{Fig:AngVelKerrToBSsLine}. Let us fix the perimetral radius, for example, $R/M = 5$. The angular velocity of the LCO at that fixed perimetral radius is increasingly smaller as $p$ increases. This implies that the quantity $R^2 \Omega^2$ (which is essentially the amplitude of the GW produced) is increasingly smaller as we add hair to the solutions. 
Additionally, if enough hair is present, $p \gtrsim 0.6$, the amplitude no longer increases monotonically. Instead, it can decrease until $R$ is close enough to the ISCO, where, if there is no extra region of UTCOs or No COs, it can increase rapidly (chirp), similarly to the extremal Kerr case. This behaviour is seen for the $p = \{0.962, 0.976, 0.981\}$ cases (solid lines) in the right panel of Fig. \ref{Fig:AmplitudeGWKerrToBSsLine}.
For solutions that have (only) an extra region of UTCOs ($0.983 \lesssim p \lesssim 0.993$), the GWs amplitude decreases further and only increases (very slightly) close to the MSCO. Lastly, for the very hairy solutions that have a region of No COs ($p \gtrsim 0.993$), the amplitude always decreases towards vanishing values as the LCO inspirals towards the BHsSH.

\begin{figure}[h!]
	\centering
	\includegraphics[scale=0.65]{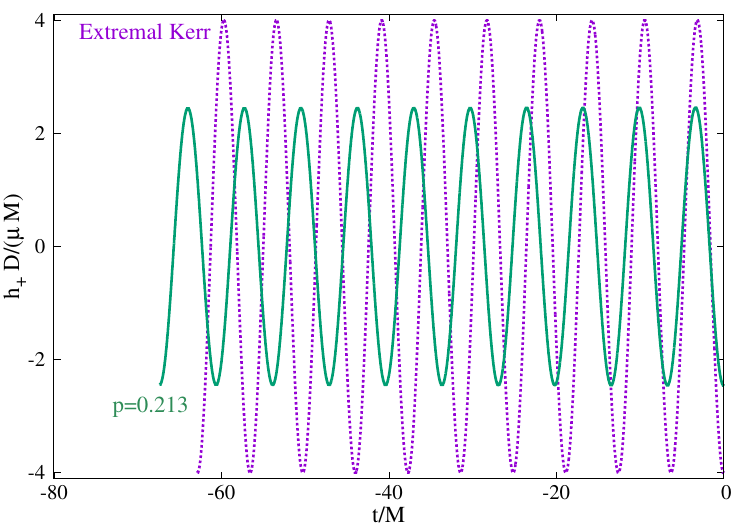}
	\includegraphics[scale=0.65]{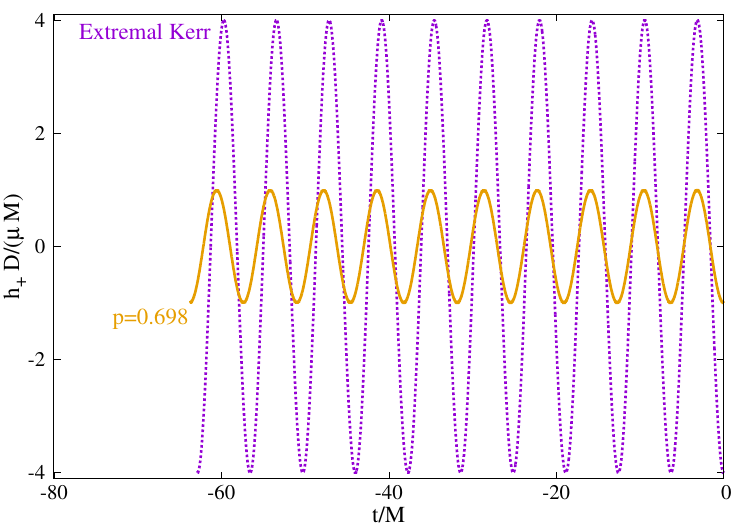}
	\includegraphics[scale=0.65]{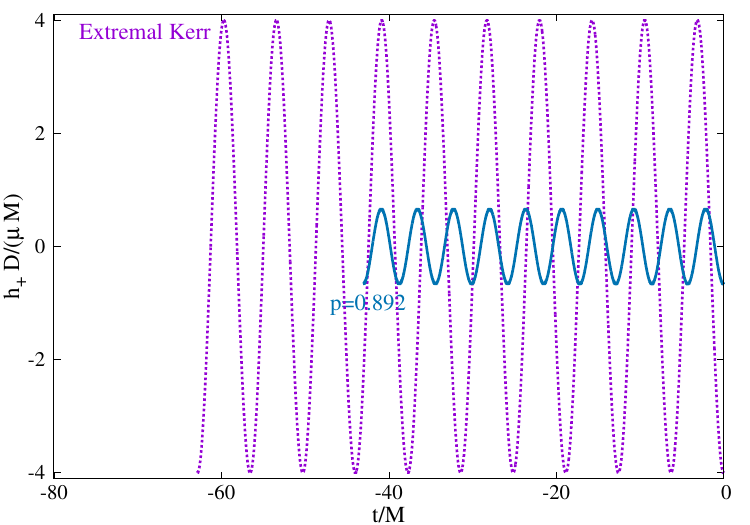}
	\includegraphics[scale=0.65]{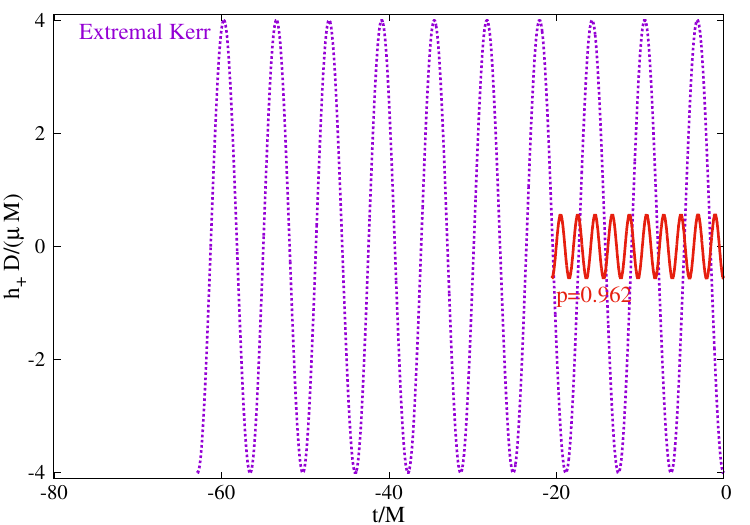}
	\caption{Waveforms generated by the last 5 revolutions of an EMRI for 4 particular cases of BHsSH in the Kerr to BS line. The waveform generated in the same setup for an extremal Kerr BH is also shown (dotted line). 
The same colour scheme as in Figs. \ref{Fig:EvolutionEMRIKerrToBSsLine} and \ref{Fig:AmplitudeGWKerrToBSsLine} was used. }
	\label{Fig:GWs_KerrToBS}
\end{figure}

For completeness, we present also here some waveforms obtained from the last 5 revolutions of the LCO around 4 BHsSH within the Kerr to BS sequence with $p=\{0.213, 0.698, 0.892, 0.962\}$. These results are presented in Fig. \ref{Fig:GWs_KerrToBS}. 
From the top panels we see that, since the endpoint of the angular velocity is slightly small than the one for the extremal Kerr BH, the LCO takes a bit more time to complete 5 revolutions, which translate into a slightly longer waveform. The reverse happens for the bottom panels. Furthermore, the amplitude of the 4 GWs shown is always smaller when compared to the extremal Kerr case, corroborating the results in Fig. \ref{Fig:AmplitudeGWKerrToBSsLine}.

We shall not present waveforms for $p \gtrsim 0.983$ because the angular velocity (and consequently the frequency of the GW) close to the end of the evolution is small, meaning that the LCO takes much more time to complete the 5 revolutions than the LCO orbiting an extremal Kerr BH. For reference, the LCO orbiting the BHsSH with $p = 0.983$ takes $t \approx 1100 M$ to complete 5 revolutions, whereas, the LCO orbiting the extremal Kerr BH takes $t \approx 60 M$. Furthermore, the amplitude of the GW produced close to the end of the evolution in the hairy solutions with $p \gtrsim 0.983$ is vanishing small compared to the amplitude of GW produced for the extremal Kerr case, as seen in Fig. \ref{Fig:AmplitudeGWKerrToBSsLine}. The disparity of scales prevents us from comparing the waveforms in the same plot as in the previous cases.

\section{Final Remarks}

Let us first summarise the main results of this work. We started by obtaining and analysing the master equation, 
Eq. (\ref{Eq:RadialEvolution}), for the evolution of an EMRI through the quadrupole hybrid formalism, wherein a LCO inspirals towards a central massive object due to the quadrupolar emission of GWs modelled by the quadrupole formula. Such a master equation was obtained by assuming that the central object is stationary, axisymmetric, asymptotically flat, described by a circular geometry and has a $\mathbb{Z}_2$ symmetry. One also assumes the LCO does not backreact on the geometry of the central object.

The master equation is informative. It determines that, far away from the central object, the LCO will inspiral towards it, as one would expect. This inspiral behaviour will terminate in one of three outcomes. 
The first is when the LCO reaches the threshold between the regions of stable and unstable timelike circular orbits. At this threshold, the variation of the perimetral radius diverges, $\dot{R} \rightarrow -\infty$, which should be interpreted as a chirp-type behaviour. 
The second outcome is when the LCO reaches a region where the function $C(R)$ changes sign. On the boundary of this region, $C(R) = 0$, which implies that $\dot{R} \rightarrow 0$, thus the evolution stalls as the LCO approaches this region. 
The same happens in the third outcome, where the LCO arrives at an orbit where the angular velocity vanishes, $\Omega_\pm = 0$, or, in other words, a SR. We further showed that a SR will always appear when $g'_{tt} = 0$, and, if the SR is in a region of STCOs, it acts as an attractor for infalling massive matter.

As an application of this general formalism, 
we analysed EMRIs around BHsSH with $j=1$, in the second part of this work.  BHsSH were first discussed in~\cite{Herdeiro:2014goa} and represent a non-linear equilibrium bound state between a boson star and a BH horizon. Equilibrium relies on a synchronization condition~\cite{Herdeiro:2014ima}, and requires both the horizon and the boson star environment to rotate - see~\cite{Herdeiro:2015gia} for an overview of these solutions. Fixing the dimensionless spin to unity yields two different sequences of solutions, one being composed of Kerr-like BHs where the maximal hairiness of the solutions is $p \sim 0.10$ -- the Kerr-like sequence --, and another composed of hairy solutions that go from the extremal Kerr BH, where $p = 0$, up to a rotating mini-boson star with $j=1$, where $p = 1$ -- the Kerr to BS sequence.

For the hairy solutions along the Kerr-like sequence,  the EMRIs do not change much from the ones around extremal Kerr BHs. However, the endpoint of the evolution has different physical parameters. The endpoint of the perimetral radius can be more than twice as large as for the extremal Kerr BH, whereas, the angular frequency endpoint, and consequently the cut-off frequency of the produced GWs, can be around one-fifth of the extremal Kerr BH value.   These differences increase monotonically with the angular frequency $\omega/m_\mu$ of the hairy BHs, but not with the hairiness of the solutions, $p$. This illustrates one cannot simply equate, heuristically, the amount of hairness with the amount non-Kerrness.  

The bottom line of this analysis is that even with a fairly small quantity of hair $p < 0.10$, the GWs produced in an EMRI can be different, in a phenomenologically significant manner, from the Kerr ones, as seen in the examples provided in Fig. \ref{Fig:AmplitudeGWKerrLikeLine}. We recall that for BHsSH formed from the growth of the superradiant instability of vacuum Kerr BHs, in the model~\eqref{action}, the bound on hairiness is $p\lesssim 0.1$~\cite{Herdeiro:2021znw}. If these BHs are supermassive they can form in an astrophysical time scale and be stable in a cosmological timescale against the next superradiant mode~\cite{Degollado:2018ypf}. This is precisely the scenario of interest for EMRIs. Thus, our analysis supports that such superradiance formed, effectively stable, BHsSH, could be potentially probed by GWs with LISA and distinguished from Kerr BHs.  This also invites a more rigorous and detailed pN analysis of the waveforms, namely including finite size effects~\cite{Pacilio:2020jza,Vaglio:2023lrd}. Considerations about the scalar field drag are also important. Even though these are almost bald solutions and the scalar field energy density small, its integrated effect in very long EMRIs may impact significantly on the evolution.

For the hairy solutions along the Kerr to BS sequence, depending on how much of the total mass is stored in the scalar field, drastically different evolutions are possible. As one increases the hairiness, the endpoint of the perimetral radius starts to decrease and the endpoint of the angular velocity (half of the frequency of the GW produced) slightly decreases at first; but increasing further the hairiness, the angular velocity endpoint increases and becomes larger than for extremal Kerr. 
For solutions with $0.94 \lesssim p \lesssim 0.983$, a non-Kerr feature starts to manifest which is the non-monotonicity of the angular velocity evolution. 
For the hairiest solutions, $p \gtrsim 0.983$, further non-Kerr features appear due to a new structure of circular orbits, the presence of new regions of UTCOs and No COs. Such new regions change drastically the evolution of the EMRI for these solutions, leading to slower evolutions and even total stagnation, as confirmed in Fig. \ref{Fig:EvolutionEMRICloseToBSsLine}.

The results in this work show that BHsSH with $j=1$ can exhibit exotic features regarding the evolution of EMRIs in the quadrupole hybrid formalism. Despite this formalism being simple and limited, it can provide valuable insight into these inspirals that can be used as a stepping stone for more accurate treatments, that we hope to pursue as future work. We also plan to develop a similar study for different hairy solutions present in different theories of Gravity, such as the case of Kerr BHs with Proca (vector) hair \cite{Herdeiro:2016tmi,Santos:2020pmh}.

\section{Acknowledgements}

This work is supported  by the  Center for Research and Development in Mathematics and Applications (CIDMA) through the Portuguese Foundation for Science and Technology (FCT -- Fundac\~ao para a Ci\^encia e a Tecnologia), references  UIDB/04106/2020 and UIDP/04106/2020.  
The authors acknowledge support  from the projects CERN/FIS-PAR/0027/2019, PTDC/FIS-AST/3041/2020,  CERN/FIS-PAR/0024/2021 and 2022.04560.PTDC.  
This work has further been supported by  the  European  Union's  Horizon  2020  research  and  innovation  (RISE) programme H2020-MSCA-RISE-2017 Grant No.~FunFiCO-777740 and by the European Horizon Europe staff exchange (SE) programme HORIZON-MSCA-2021-SE-01 Grant No.~NewFunFiCO-101086251. Computations have been performed at the Argus and Blafis cluster at the U. Aveiro.

\bibliographystyle{ieeetr}
\bibliography{Bibliography}

\end{document}